\documentclass[a4paper,11pt]{article}
\usepackage{latexsym}
\usepackage{amssymb}
\usepackage{amsfonts}
\usepackage{amsmath}
\usepackage{indentfirst}
\usepackage[dvips]{graphicx}
\usepackage{epsfig}
\newcommand{\cvd}{\hfill $\blacksquare$\bigskip}

\newtheorem{theorem}{Theorem}[section]

\date{}

\author{S. Bilotta\thanks{Dipartimento di Sistemi e Informatica, Universit\`a degli Studi di Firenze, Viale
 G.B. Morgagni 65, 50134 Firenze, Italy. {\newline
 \tt \ bilotta@dsi.unifi.it,\quad
donatella.merlini@unifi.it,\quad elisa@dsi.unifi.it,\quad
pinzani@dsi.unifi.it}} \and D. Merlini$^*$\and E. Pergola$^*$\and
R. Pinzani$^*$}

\title{Binary words avoiding a pattern and marked succession rule}

\begin{document}

\maketitle

\begin{abstract}
In this paper we study the enumeration and the construction of
particular binary words avoiding the pattern $1^{j+1}0^j$. By
means of the theory of Riordan arrays, we solve the enumeration
problem and we give a particular succession rule, called
\emph{jumping and marked succession rule}, which describes the
growth of such words according to their number of ones. Moreover,
the problem of associating a word to a path in the generating tree
obtained by the succession rule is solved by introducing an
algorithm which constructs all binary words and then kills those
containing the forbidden pattern.
\end{abstract}

\section{Introduction}
Binary words avoiding a given pattern
$\mathfrak{p}=p_0...p_{h-1}\in \{0,1\}^h$ constitute  a regular
language and can be enumerated in terms of the number of bits $1$
and $0$ by using classical results (see, e.g., \cite{9,10,15}).
Recently, in \cite{2,12}, this subject has been studied in
relation to the theory of Riordan arrays. The concept of
\emph{Riordan array} has been introduced in 1991 by Shapiro, Getu,
Woan and Woodson \cite{16}, with the aim of defining a class of
infinite lower triangular arrays with properties analogous to
those of the Pascal's triangle.

Riordan arrays have been studied in relation to \emph{succession
rules} and \emph{generating trees} associated to a certain
combinatorial class, according to some enumerative parameter. In
particular, we use an algebraic approach (see \cite{12}) to study
the connection between proper Riordan arrays and succession rules
describing the growth, according to the number of ones, of
particular binary words avoiding some fixed pattern
$\mathfrak{p}$.

In Section 2, we give some basic definitions and notation related
to the notions of succession rule and generating tree. In
particular, we introduce the concept of \emph{jumping and marked
succession rules} (see \cite{7,8}) which are succession rules
acting on the combinatorial objects of a class and producing sons
at different levels where appear marked or non-marked labels.

In Section 3, we give necessary and sufficient conditions for the
number of words counted according to the number of their zeroes
and ones to be related to proper Riordan arrays.

In Section 4, by means of the theory of Riordan arrays we solve
algebraically the enumeration problem, according to the number of
ones. This approach enables us to obtain a jumping and marked
succession rule describing the growth of such words. In particular
we show that, when the forbidden pattern has a particular shape,
then each row of the related Riordan array corresponds to a level
of the generating tree which generates all the binary words
avoiding the pattern.

We will show that it is not possible to associate to a word a path
in the generating tree obtained by the succession rule. The
problem is solved in Section 5, where we introduce an algorithm
for constructing all binary words having a fixed number of ones
and excluding those containing the forbidden pattern
$\mathfrak{p}$.

\section{Basic definitions and notations}
A \emph{succession rule} $\Omega$ is a system constituted by an
\emph{axiom} $(a)$, with $a \in \mathbb{N}$, and a set of
\emph{productions} of the form:
\begin{displaymath}
(k)\rightsquigarrow (e_1(k))(e_2(k))\ldots(e_k(k)), \ \ \ \ \ k
\in \mathbb{N}, \ e_i : \mathbb{N} \rightarrow \mathbb{N}.
\end{displaymath}

A production constructs, for any given label $(k)$, its
\emph{successors} $(e_1(k)),(e_2(k)),\ldots,(e_k(k))$. In most of
the cases, for a succession rule $\Omega$, we use the more compact
notation:
\begin{equation}
\label{uno} \left\{
\begin{array}{cl}
 (a) & \\
 (k) & \rightsquigarrow (e_1(k))(e_2(k))\ldots(e_k(k))
\end{array}
\right.
\end{equation}

The rule $\Omega$ can be represented by means of a
\emph{generating tree}, that is a rooted tree whose vertices are
the labels of $\Omega$; where $(a)$ is the label of the root and
each node labelled $(k)$ produces $k$ sons labelled
$(e_1(k)),(e_2(k)),\ldots,(e_k(k))$, respectively. As usual, the
root lies at level 0, and a node lies at level $n$ if its parent
lies at level $n-1$. If a succession rule describes the growth of
a class of combinatorial objects, then a given object can be coded
by the sequence of labels met from the root of the generating tree
to the object itself. We refer to \cite{4} for further details and examples.

The concept of succession rule was introduced in \cite{6} by Chung
et al. to study reduced Baxter permutations, and was later applied
to the enumeration of permutations with forbidden subsequences
\cite{5,18}.

We remark that, from the above definition, a node labelled $(k)$
has precisely $k$ sons. In \cite{1}, a succession rule having this
property is said to be \emph{consistent}. However, we can also
consider succession rules, introduced in \cite{7}, in which the
value of
a label does not necessarily represent the number of its sons,
and this will be frequently done in the sequel.

Regular succession rules are not sufficient to handle all
enumeration problems and so we consider a slight generalization
called \emph{jumping succession rule}. Roughly speaking, the idea
is to consider a set of succession rules acting on the objects of
a class and producing sons at different levels.

The usual notation to indicate a jumping succession rule
is the following:
\begin{equation}
\label{due} \left\{
\begin{array}{cl}
 (a) & \\
 (k) & \stackrel{1}{\rightsquigarrow} (e_1(k))(e_2(k))\ldots(e_k(k))\\
 (k) & \stackrel{j}{\rightsquigarrow} (d_1(k))(d_2(k))\ldots(d_k(k))
\end{array}
\right.
\end{equation}

\begin{figure}[htb]
\begin{center}
\epsfig{file=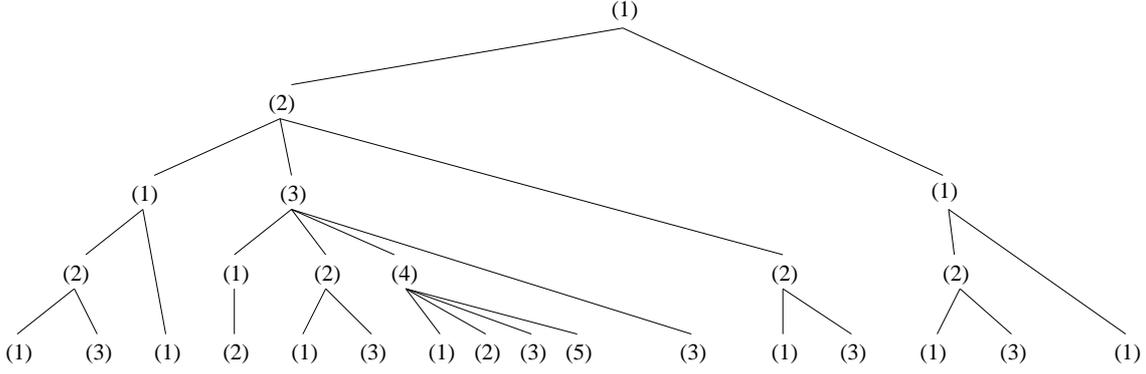,width=5.9in,clip=} \caption{\small{Four
levels of the generating tree associated with the succession rule
(\ref{jump})} \label{jp}}\vspace{-15pt}
\end{center}
\end{figure}

The generating tree associated with (\ref{due}) has the property
that each node labelled $(k)$ lying at level $n$ produces two sets
of sons, the first set at level $n+1$ and having labels
$(e_1(k)),(e_2(k)),\ldots,(e_k(k))$ respectively and the second
one at level $n+j$, with $j>1$, and having labels
$(d_1(k)),(d_2(k)),\ldots,(d_k(k))$ respectively.
For example, the jumping succession rule (\ref{jump}) counts the number of
\emph{2-generalized Motzkin paths}
and Figure \ref{jp} shows some levels of the associated generating tree.
For more details about these topics, see \cite{8}.
\begin{equation}\label{jump}
\left\{ \begin{array}{lll} (1)&
\\ (k)&\stackrel{1}{\rightsquigarrow} (1)(2)\cdots (k-1)(k+1)
\\ (k)&\stackrel{2}{\rightsquigarrow} (k)
\end{array}\right.
\end{equation}

Another generalization is introduced in \cite{13}, where the
authors deal with \emph{marked succession rules}. In this case the
labels appearing in a succession rule can be marked or not,
therefore \emph{marked} are considered together with usual labels.
In this way a generating tree can support negative values if we
consider a node labelled $(\overline{k})$ as opposed to a node
labelled $(k)$ lying on the same level.

A \emph{marked generating tree} is a rooted labelled
tree where appear marked or non-marked labels according to the
corresponding succession rule. The main property is that, on the
same level, marked labels kill or annihilate the non-marked ones
with the same label value, in particular the enumeration of the
combinatorial objects in a class is the difference between the
number of non-marked and marked
labels lying on a given level.

For any label $(k)$, we introduce the following notation for
generating tree specifications:
\begin{itemize}
\item[] $(\overline{\overline{k}})=(k)$; \item[] $(k)^n =
\underbrace{(k)\ldots(k)}_{n}, \ \ n > 0.$
\end{itemize}

Each succession rule (\ref{uno}) can be trivially rewritten as
(\ref{tre})
\begin{equation}
\label{tre} \left\{
\begin{array}{ll}
 (a) & \\
 (k) & \rightsquigarrow (e_1(k))(e_2(k))\ldots(e_k(k))(k)\\
 (k) & \rightsquigarrow (\overline{k})
\end{array}
\right.
\end{equation}

\begin{figure}[htb]
\begin{center}
\epsfig{file=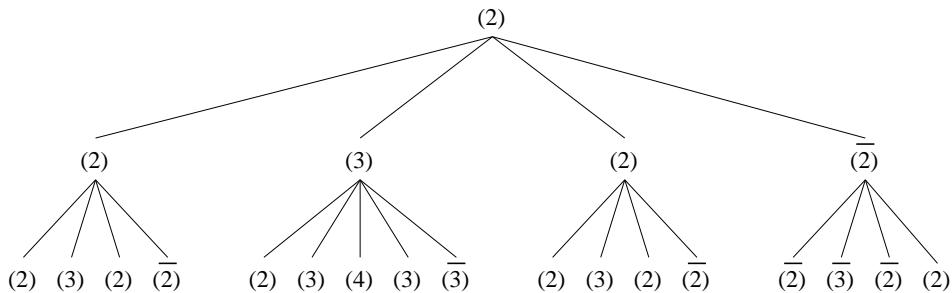,width=4.9in,clip=} \caption{\small{Three
levels of the generating tree associated with the succession rule
(\ref{quattro})} \label{exmar}}\vspace{-15pt}
\end{center}
\end{figure}

For example, the classical succession rule for Catalan
numbers can be rewritten in the form (\ref{quattro}) and Figure
\ref{exmar} shows some levels of the associated generating tree.
\begin{equation}
\label{quattro} \left\{
\begin{array}{ll}
 (2) & \\
 (k) & \rightsquigarrow (2)(3)\ldots(k)(k+1)(k)\\
 (k) & \rightsquigarrow(\overline{k})
\end{array}
\right.
\end{equation}

The concept of marked labels has been implicity used for the first
time in \cite{14}, then in \cite{7} in relation with the
introduction of the signed ECO-systems. In Section 4, we show how
marked succession rules appear in the enumeration of a class of
particular binary words according to the number of ones. Let $F
\subset \{0,1\}^*$ be the class of binary words $w$ such that
$|w|_0 \leq |w|_1$ for any $w \in F$, $|w|_0$ and $|w|_1$ are the
number of zeroes and ones in $w$, respectively.

In this paper we are interested in studying the subclass
$F^{[\mathfrak{p}]}$ of $F$ of binary words excluding a given
pattern $\mathfrak{p}=p_0 \ldots p_{h-1} \in \{0,1\}^h$, i.e. the
word $w \in F^{[\mathfrak{p}]}$ that does not admit a sequence of
consecutive indices $i,i+1,\ldots,i+h-1$ such that $w_i w_{i+1}
\ldots w_{i+h-1} = p_0 p_1 \ldots p_{h-1}$. Each word $w \in F$
can be naturally represented as a lattice path on the Cartesian
plane by associating a \emph{rise step}, defined by $(1,1)$ and
denoted by $x$, to each 1's in $F$, and a \emph{fall step},
defined by $(1,-1)$ and denoted by $\overline{x}$, to each 0's in
$F$. From now on, we refer interchangeably to words or their
graphical representations on the Cartesian plane, that is paths.
\section{Binary words avoiding a pattern and Riordan arrays}
In this section, we establish necessary and sufficient conditions
for the number of words counted according to the number of zeroes
and ones to be related to proper Riordan arrays.

This problem is interesting in the context of the Riordan arrays
theory because the matrices arising there are naturally defined by
recurrence relations following the characterization given in
\cite{11} (see formula (\ref{recAA}) below). In particular, if
$F_{n,k}^{[\mathfrak{p}]}$ denotes the number of words excluding
the pattern and having $n$ bits $1$ and $k$ bits $0,$ then by
using the results in \cite{2} we have
\begin{equation}\label{F_F}
F^{[\mathfrak{p}]}(x,y)=\sum_{n,k \geq
0}F_{n,k}^{[\mathfrak{p}]}x^ny^k ={C^{[\mathfrak{p}]}(x,y) \over
(1-x-y)C^{[\mathfrak{p}]}(x,y) +x^{|\mathfrak{p}|_1}
y^{|\mathfrak{p}|_0}},
\end{equation}
where $|\mathfrak{p}|_1$ and  $|\mathfrak{p}|_0$ correspond to the
number of ones and zeroes in the pattern and
$C^{[\mathfrak{p}]}(x,y)$ is the autocorrelation polynomial with
 coefficients  given by the  autocorrelation vector  (see also \cite{9,10,15}).
For a given $\mathfrak{p}$, this vector of bits $c=(c_0,\ldots
,c_{h-1})$ can be defined in terms of Iverson's bracket notation
(for a predicate $P$, the expression $[\![P]\!]$ has value 1 if
$P$ is true and 0 otherwise) as follows: $c_i=[\![p_0p_1\cdots
p_{h-1-i}=p_{i}p_{i+1}\cdots p_{h-1}]\!]$. In other words, the bit
$c_i$ is determined by shifting $\mathfrak{p}$ right by $i$
positions and setting  $c_i=1$ iff the remaining letters match the
original. For example, when $\mathfrak{p}= 101010$ the
autocorrelation vector is $c=(1,0,1,0,1,0),$ as illustrated in
Table \ref{auto}, and $C^{[\mathfrak{p}]}(x,y)=1+xy+x^2y^2,$ that
is, we mark with $x^jy^i$ the tails of the pattern with $j$ bits
$1,$ $i$ bits $0$ and $c_{j+i}=1.$ Therefore, in this case we
have:
$$F^{[\mathfrak{p}]}(x,y)={1+xy+x^2y^2 \over (1-x-y)(1+xy+x^2y^2)+x^3y^3}. $$
\begin{table}[htb]
\begin{center}
\begin{tabular}{cccccc|ccccccc}
  1 & 0 & 1 & 0& 1 & 0 &  \multicolumn{7}{|c}{Tails}  \\
  \hline
  1 & 0 & 1 & 0 & 1 & 0 &   &   &   &   &   &   & 1 \\
    & 1 & 0 & 1 & 0 & 1 & 0 &   &   &   &   &   & 0 \\
    &   & 1 & 0 & 1 & 0 & 1 & 0 &   &   &   &   & 1 \\
    &   &   & 1 & 0 & 1 & 0 & 1 & 0 &   &   &   & 0 \\
    &   &   &   & 1 & 0 & 1 & 0 & 1 & 0 &   &   & 1 \\
    &   &   &   &   & 1 & 0 & 1 & 0 & 1 & 0 &   & 0 \\
\end{tabular}
\end{center}
\caption{\label{auto}The autocorrelation vector for $\mathfrak{p}=
101010$}
\end{table}
As another example, when $\mathfrak{p}= 11100$ then
$C^{[\mathfrak{p}]}(x,y)=1$ and
$F^{[\mathfrak{p}]}(x,y)=1/(1-x-y+x^3y^2). $

\begin{table}[htb]
$$
\begin{array}{c|cccccccc}
n/k  & 0 & 1 & 2 & 3 & 4 &5 &6 &7  \\ \cline{1-9}
0 & 1 & 1 & 1 & 1& 1 &  1 & 1 & 1\\
1 & 1 & 2 & 3 & 4 & 5 & 6 & 7 & 8 \\
2 & 1 & 3 & 6 & 10 & 15 & 21 & 28 & 36\\
3 & 1 & 4 & 9 & 18 & 32 & 52& 79 & 114\\
4 & 1& 5 & 13 & 29 & 58 & 106& 180 & 288\\
5 & 1 & 6 & 18 & 44 & 96 & 192 & 357 & 624\\
6 & 1 & 7 & 24 & 64 & 151 & 325& 650 & 1222\\
7 & 1 & 8 & 31 & 90 & 228 &524& 1116 &2232\\
\end{array}
$$
\caption{\label{Fprimo}The  matrix  ${\cal{F}^{[\mathfrak{p}]}}$
for $\mathfrak{p}=11100$}
\end{table}

\begin{table}[htb]
$$
\begin{array}{c|cccccccc}
n/k  & 0 & 1 & 2 & 3 & 4 &5 &6 &7  \\ \cline{1-9}
0 & 1 & & & & & & &\\
1 & 2 & 1 & & &  & & &\\
2 & 6& 3 & 1 & & & &  &\\
3 & 18 & 9 & 4 & 1 & & &  &\\
4 & 58 & 29 & 13 & 5 & 1& & & \\
5 & 192 & 96 & 44 & 18 & 6 &1 & &\\
6 & 650 & 325 & 151 & 64 & 24 & 7& 1 &\\
7 & 2232 & 1116&  524 &228 & 90 & 31 &  8 &1\\
\end{array}
$$
\caption{\label{Rprimoa}The  triangle ${\cal{R}^{[\mathfrak{p}]}}$
for $\mathfrak{p}=11100$}
\end{table}

\begin{table}[htb]
$$
\begin{array}{c|cccccccc}
n/k  & 0 & 1 & 2 & 3 & 4 &5 &6 &7  \\ \cline{1-9}
0 & 1 & & & & & & &\\
1 & 2 & 1 & & &  & & &\\
2 & 6& 3 & 1 & & & &  &\\
3 & 18 & 10 & 4 & 1 & & &  &\\
4 & 58 & 32 & 15 & 5 & 1& & & \\
5 & 192 & 106 & 52 & 21 & 6 &1 & &\\
6 & 650 & 357 & 180 & 79 & 28 & 7& 1 &\\
7 & 2232 & 1222 &  624 &288 & 114 & 36 &  8 &1\\
\end{array}
$$
\caption{\label{Rprimob}The  triangle
${\cal{R}^{[\bar{\mathfrak{p}]}}}$ for $\bar{\mathfrak{p}}=00011$}
\end{table}
In order to study the binary words avoiding a pattern in terms of
Riordan arrays, we consider the array
${\cal{R}^{[\mathfrak{p}]}}=(R_{n, k}^{[\mathfrak{p}]})$ given by
the lower triangular part of the array
${\cal{F}^{[\mathfrak{p}]}}=(F_{n,k}^{[\mathfrak{p}]}),$ that is,
$R_{n,k}^{[\mathfrak{p}]}=F_{n,n-k}^{[\mathfrak{p}]}$ with $k\leq
n.$  More precisely, $R_{n,k}^{[\mathfrak{p}]}$ counts the number
of words avoiding $\mathfrak{p}$ and having length $2n-k,$  $n$
bits one and $n-k$  bits zero. Given a pattern
$\mathfrak{p}=p_{0}\ldots p_{h-1}\in\{0,1\}^h$, let
$\bar{\mathfrak{p}}=\bar{p}_{0}\ldots\bar{p}_{h-1}$ be the pattern
with
 $\bar{p}_{i}=1-p_{i},\forall i=0,\cdots,h-1$.
We obviously have
$R_{n,k}^{[\bar{\mathfrak{p}}]}=F_{n,n-k}^{[\bar{\mathfrak{p}}]}=F_{n-k,n}^{[\mathfrak{p}]},$
therefore, the matrices ${\cal{R}^{[\mathfrak{p}]}}$ and
${\cal{R}^{[\bar{\mathfrak{p}]}}}$ represent the lower and upper
triangular part of the array ${\cal{F}^{[\mathfrak{p}]}},$
respectively. Moreover, we have
$R_{n,0}^{[\mathfrak{p}]}=R_{n,0}^{[\bar{\mathfrak{p}}]}=F_{n,n}^{[\mathfrak{p}]},$
$\forall n \in \mathbb{N},$ that is,  columns zero of
${\cal{R}^{[\mathfrak{p}]}}$ and
${\cal{R}^{[\bar{\mathfrak{p}]}}}$ correspond to the main diagonal
of ${\cal{F}^{[\mathfrak{p}]}}.$ Tables \ref{Fprimo},
\ref{Rprimoa} and \ref{Rprimob} illustrate some rows for the
matrices ${\cal{F}^{[\mathfrak{p}]}},$
${\cal{R}^{[\mathfrak{p}]}}$ and
${\cal{R}^{[\bar{\mathfrak{p}]}}}$ when $\mathfrak{p}=11100.$

We briefly recall that a Riordan array is an infinite lower
triangular array  $(d_{n,k} )_{n,k \in \mathbb{N}},$ defined by a
pair of formal power series $(d(t),h(t)),$ such that $d(0)\neq 0,
h(0)=0, h^\prime(0)\neq0$ and the generic element $d_{n,k}$ is the
$n$-th coefficient in the series
  $d(t)h(t)^k,$ i.e.:
$$ d_{n,k}=[t^n]d(t)h(t)^k, \qquad n,k \geq 0. $$
From this definition we have  $d_{n,k}=0$ for $k>n.$ An
alternative definition is in terms of the so-called $A$-sequence
and $Z$-sequence, with generating functions $A(t)$ and $Z(t)$
satisfying the relations:
 $$ h(t)=tA(h(t)), \quad d(t)={d_0\over 1-tZ(h(t))} \quad \mbox{with} \quad d_0=d(0).$$

 In other words,  Riordan arrays correspond to matrices where each element $d_{n,k}$ is described
 by a linear combination of the elements in the previous row, starting from the previous column,
 with coefficients in $A$:
 \begin{equation}
 \label{RecA}
 d_{n+1,k+1}=a_0d_{n,k}+a_1d_{n,k+1}+a_2d_{n,k+2}+\cdots
 \end{equation}

Another characterization (see \cite{11}) states that a lower
triangular array $(d_{n,k})_{n,k \in \mathbb{N}}$ is Riordan if
and only if there exists another array $(\alpha_{i,j})_{i,j \in
\mathbb{N}}$, with $\alpha_{0,0}\neq 0$,
 and a sequence $(\rho_j)_{j\in \mathbb{N}}$ such that:
\begin{equation}
\label{recAA} d_{n+1,k+1}={\underset{i\geq0}{ \displaystyle\sum
}}{\underset{j\geq0}{ \displaystyle\sum }}\alpha_{i,j}d_{n-i,k+j}+
\underset{j\geq 0}{\displaystyle\sum}\rho_jd_{n+1,k+j+2}.
\end{equation}
 Matrix $(\alpha_{i,j})_{i,j\in \mathbb{N}}$  is
called the $A$-matrix of the Riordan array. If  $P^{[0]}(t),
P^{[1]}(t), P^{[2]}(t),\ldots$ denote the generating functions of
rows $0,1,2,\ldots$ in the $A$-matrix, i.e.:
$$P^{[i]}(t)=\alpha_{i,0}+\alpha_{i,1}t+\alpha_{i,2}t^2+\alpha_{i,3}t^3+\ldots$$
 and $Q(t)$ is the generating function for the
sequence $(\rho_j)_{j\in \mathbb{N}}$, then we have:
\begin{equation}\label{hAA}
\dfrac{h(t)}{t}=\underset{i\geq
0}{\displaystyle\sum}t^iP^{[i]}(h(t))+\dfrac{h(t)^{2}}{t}Q(h(t)),
\end{equation}
\begin{equation}\label{At}
A(t)=\underset{i\geq
0}{\displaystyle\sum}t^iA(t)^{-i}P^{[i]}(t)+tA(t)Q(t).
\end{equation}

The  theory of Riordan arrays  and  the proofs of their properties
can be found in \cite{11}. The Riordan arrays which arise in the
context of pattern avoidance (see \cite{2,12}) have the nice
property to be defined by a quite simple recurrence relation
following the characterization (\ref{recAA}), while the relation
induced by the $A$-sequence is, in general, more complex. From a
combinatorial point of view,  this means that it is very
challenging to find a construction allowing to build objects of
size $n + 1$ from objects of size $n$. Instead, the existence of a
simple $A$-matrix corresponds to a possible construction from
objects of different sizes less than $n + 1$. On the other hand,
as we will see in Section \ref{Riordan}, the recurrence following
characterization (\ref{recAA}) contains negative coefficients and
therefore gives rise to interesting non trivial combinatorial
problems.

In this paper we examine in particular the family of patterns
$\mathfrak{p}=1^{j+1}0^j$ and show that the corresponding
recurrence relation can be combinatorially interpreted. To this
purpose, we translate the recurrence into a succession rule, as it
is typically done from problems related to Riordan arrays (see,
e.g., \cite{3,14}), and give a construction for the class of
binary words avoiding the pattern $\mathfrak{p}.$

\section{The Riordan array for the pattern $\mathfrak{p}=1^{j+1}0^j$}
\label{Riordan} Let us consider the family of patterns
$\mathfrak{p}=1^{j+1}0^j$ and let $F_{n,k}^{[\mathfrak{p}]}$
denote the number of words excluding the pattern and having $n$
bits $1$ and $k$ bits $0;$ from (\ref{F_F}) we have
\begin{equation}
\label{Fgen} F^{[\mathfrak{p}]}(x,y)=\sum_{n,k \geq
0}F_{n,k}^{[\mathfrak{p}]}x^ny^k={ 1 \over 1-x-y+x^{j+1}y^j}.
\end{equation}

Now, let $R_{n,k}^{[\mathfrak{p}]}$ count the number of words
avoiding $\mathfrak{p}$ and having  $n$ bits one and $n-k$  bits
zero. Obviously we have
$R_{n,k}^{[\mathfrak{p}]}=F_{n,n-k}^{[\mathfrak{p}]}$ with $k\leq
n.$ By extracting the coefficients from  (\ref{Fgen}) we have:
$$[x^{n+1}y^{k+1}](1-x-y+x^{j+1}y^j)F^{[\mathfrak{p}]}(x,y)=F_{n+1,k+1}^{[\mathfrak{p}]}-F_{n,k+1}^{[\mathfrak{p}]}-
F_{n+1,k}^{[\mathfrak{p}]}+ F_{n-j,k+1-j}^{[\mathfrak{p}]}=0$$ and
therefore:
\begin{equation}
\label{recRA} R_{n+1, k+1}^{[\mathfrak{p}]}=R_{n,
k}^{[\mathfrak{p}]}+R_{n+1, k+2}^{[\mathfrak{p}]}-R_{n-j,
k}^{[\mathfrak{p}]}.
\end{equation}

This is a recurrence relation of type (\ref{recAA}) and therefore
${\cal{R}^{[\mathfrak{p}]}}=(R_{n, k}^{[\mathfrak{p}]})$ is a
Riordan array. In particular, the coefficients of the relation
 correspond
to $P^{[j]}(t)=-1,$ $P^{[0]}(t)=1,$ and $Q(t)=1,$ therefore we
have
$${h^{[\mathfrak{p}]}(t) \over t}=\sum_{i \geq 0}t^i P^{[i]}(h^{[\mathfrak{p}]}(t))+
{h^{[\mathfrak{p}]}(t)^2 \over t}Q(h(t))=1-t^j+
{h^{[\mathfrak{p}]}(t)^2 \over t}$$ that is,
$$h^{[\mathfrak{p}]}(t)^2-h^{[\mathfrak{p}]}(t)+t-t^{j+1}=0, \quad h^{[\mathfrak{p}]}(t)=
{1-\sqrt{1-4t+4t^{j+1}} \over 2}. $$

We explicitly observe that
from formula (\ref{At}) the generating function $A(t)$  of the
$A$-sequence is the solution of a $j+1$ degree equation
$(1-t)A(t)^{j+1}-A(t)^j+t^j=0.$  For example, when
$\mathfrak{p}=11100$ by developing into series we find:
$$A(t)= 1+t+2t^3-t^4+7t^5-12t^6+38t^7-99t^8+281t^9+O(t^{10})$$
and this result excludes that there might exist a "simple"
dependence of the elements in row $n+1$ from the elements in row
$n$. For what concerns $d^{[\mathfrak{p}]}(t),$ we simply use  the
Cauchy formula for finding the main diagonal of matrix
${\cal{F}^{[\mathfrak{p}]}}$
 (see, e.g., \cite[Cap. 6, p. 182]{17}):
$$ d^{[\mathfrak{p}]}(t)=[x^0]F^{[\mathfrak{p}]}(x,\dfrac{t}{x})=\dfrac{1}{2\pi
i}\displaystyle\oint{F^{[\mathfrak{p}]}(x,\dfrac{t}{x})\dfrac{dx}{x}}.
$$
We have:
$${1 \over x}F^{[\mathfrak{p}]}(x,\dfrac{t}{x})={-1 \over x^2(1-t^j)-x+t}$$
and in order to compute the integral, it is necessary to find the
singularities $x(t)$ such that $x(t)\rightarrow 0$ with
$t\rightarrow 0$ and apply the Residue theorem. In this case  the
right singularity is:
$$x(t)={1-\sqrt{1-4t(1-t^j)} \over 2(1-t^j)} $$ and finally we have:
$$d^{[\mathfrak{p}]}(t)=\lim _{x \rightarrow x(t)} {-1 \over x^2(1-t^j)-x+t}(x-x(t))={1 \over \sqrt{1-4t+4t^{j+1}}}. $$

Observe also that:
$${d^{[\mathfrak{p}]}(t) -1 \over d^{[\mathfrak{p}]}(t)h^{[\mathfrak{p}]}(t) }=2 $$
and therefore $R_{n+1, 0}^{[\mathfrak{p}]}=2R_{n+1,
1}^{[\mathfrak{p}]}. $ Recurrence (\ref{recRA}) is quite simple,
however, the presence of negative coefficients leads to a possible
non trivial combinatorial interpretation. In order to study this
problem we proceed as follows. The dependence of  $R_{n+1,
k+1}^{[\mathfrak{p}]}$ from the same row $n+1$ can be simply
eliminated and we have:
$$R_{n+1, k+1}^{[\mathfrak{p}]}=R_{n, k}^{[\mathfrak{p}]}-R_{n-j, k}^{[\mathfrak{p}]}+R_{n+1, k+2}^{[\mathfrak{p}]}=$$
$$ =R_{n, k}^{[\mathfrak{p}]}-R_{n-j, k}^{[\mathfrak{p}]}+R_{n, k+1}^{[\mathfrak{p}]}-
R_{n-j, k+1}^{[\mathfrak{p}]}+R_{n+1,
k+3}^{[\mathfrak{p}]}=\cdots=$$
\begin{equation}
\label{reck} =(R_{n, k}^{[\mathfrak{p}]}+R_{n,
k+1}^{[\mathfrak{p}]}+R_{n, k+2}^{[\mathfrak{p}]} + \cdots)-
(R_{n-j, k}^{[\mathfrak{p}]}+R_{n-j, k+1}^{[\mathfrak{p}]}+
R_{n-j, k+2}^{[\mathfrak{p}]}+\cdots)
\end{equation}
Similarly we have:
\begin{equation}
\label{rec0} R_{n+1, 0}^{[\mathfrak{p}]}= 2(R_{n,
0}^{[\mathfrak{p}]}+R_{n, 1}^{[\mathfrak{p}]}+R_{n,
2}^{[\mathfrak{p}]} + \cdots)- 2(R_{n-j,
0}^{[\mathfrak{p}]}+R_{n-j, 1}^{[\mathfrak{p}]}+ R_{n-j,
2}^{[\mathfrak{p}]}+\cdots)
\end{equation}

Finally, by using the results in \cite{2,3}, recurrences
(\ref{reck}) and (\ref{rec0}) translate into the following
succession rule:
\begin{equation}
\label{Rule} \left\{
\begin{array}{cl}
 (0) & \\
 (k) & \stackrel{1}{\rightsquigarrow} (0_1)(0_2) (1) \cdots (k+1)\\
 (k) & \stackrel{j+1}{\rightsquigarrow} (\overline{0_1})(\overline{0_2}) (\overline{1}) \cdots (\overline{k+1})
\end{array}
\right.
\end{equation}

This rule can be represented as a tree having its root labelled
$(0)$ and where each node with label $(k)$ at a given level $n$
has $k+3$ sons at level $n+1$ labelled
$(0_1),(0_2),(1),\cdots,(k+1)$ and $k+3$ sons at level $n+j+1$
with labels
$(\overline{0_1}),(\overline{0_2}),(\overline{1}),\cdots,(\overline{k+1})$
(this kind of trees are called \textit{level generating trees} in
\cite{3}). If we denote by $d_{n,k}$ the number of nodes having
label $k$ at level $n$ in the tree  and count as negative the
marked nodes then we obtain matrix
${\cal{R}^{[\mathfrak{p}]}}=(R_{n,k}^{[\mathfrak{p}]})_{n,k\in\mathbb{N}},$
that is, ${\cal{R}^{[\mathfrak{p}]}}$ corresponds to the matrix
associated to the rule (\ref{Rule}). The relations between Riordan
arrays and succession rules has been widely studied and we refer
the reader to \cite{3,13,14} for more details. We just conclude
this section by observing that by using the results in \cite{2,12}
it can  be proved that the matrix
${\cal{R}^{[\bar{\mathfrak{p}}]}}$ corresponding to the  pattern
$\bar{\mathfrak{p}}=0^{j+1}1^j$ is also a Riordan array.

\section{A construction for the class $F^{[\mathfrak{p}]}$}
In this section we define an algorithm which associates a lattice
path in $F^{[\mathfrak{p}]}$, where
$\mathfrak{p}=x^{j+1}\overline{x}^j=1^{j+1}0^j$, to a sequence of
labels obtained by means of the succession rule (\ref{Rule}). This
give a construction for the set $F^{[\mathfrak{p}]}$ according to
the number of rise steps or equivalently the number of ones.

\begin{figure}[!htb]
\begin{center}
\epsfig{file=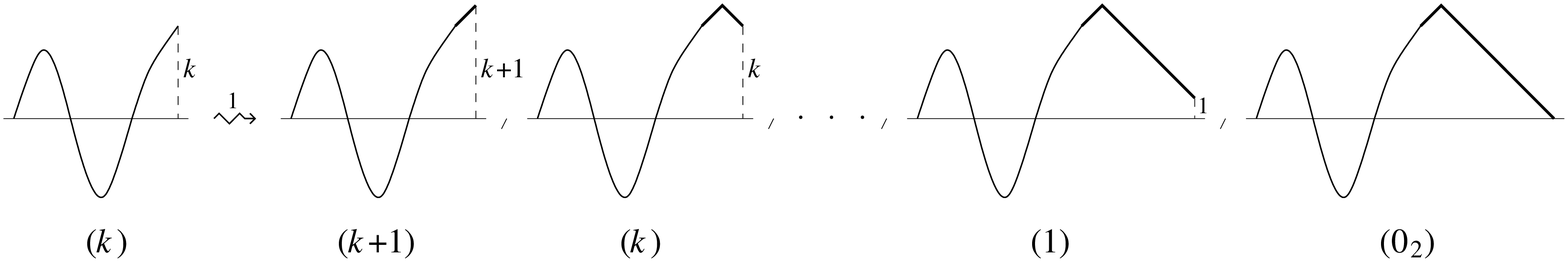,width=5.8in,clip=} \caption{\small{The
mapping associated to $(k) \stackrel{1}{\rightsquigarrow}
(0_2)(1)\ldots(k+1)$ of (\ref{Rule})} \label{sopra}}\vspace{-15pt}
\end{center}
\end{figure}

The axiom $(0)$ is associated to the empty path $\varepsilon$.\\
A lattice path $\omega \in F$, with $n$ rise steps and such that
its endpoint has ordinate $k$, provides $k+3$ lattice paths with
$n+1$ rise steps, according to the first production of
(\ref{Rule}) having $0,0,1,\ldots,k+1$ as endpoint ordinate,
respectively. The last $k+2$ labels are obtained by adding to
$\omega$ a sequence of steps consisting of one rise step followed
by $k+1-h$, $0 \leq h \leq k+1$, fall steps (see Figure
\ref{sopra}). Each lattice path so obtained has the property that
its rightmost suffix beginning from the $x$-axis, either remains
strictly above the $x$-axis itself or ends on the $x$-axis by a
fall step. Note that in this way the paths ending on the $x$-axis
and having a rise step as last step are never obtained. These
paths are bound to the label $(0_1)$ of the first production
in (\ref{Rule}) and the way to obtain them will be described later
in this section.

We define a \emph{marked forbidden pattern} $\mathfrak{p}$ as a
pattern $\mathfrak{p}=x^{j+1}\overline{x}^j$ whose steps cannot be
divided, they must lie always in that defined sequence. Therefore,
a cut operation is not possible within a marked forbidden pattern
$\mathfrak{p}$. We denote a marked forbidden pattern by marking
its peak. We say that a point is strictly contained in a marked
forbidden pattern if it is between two steps of the pattern
itself.

\begin{figure}[htb]
\begin{center}
\epsfig{file=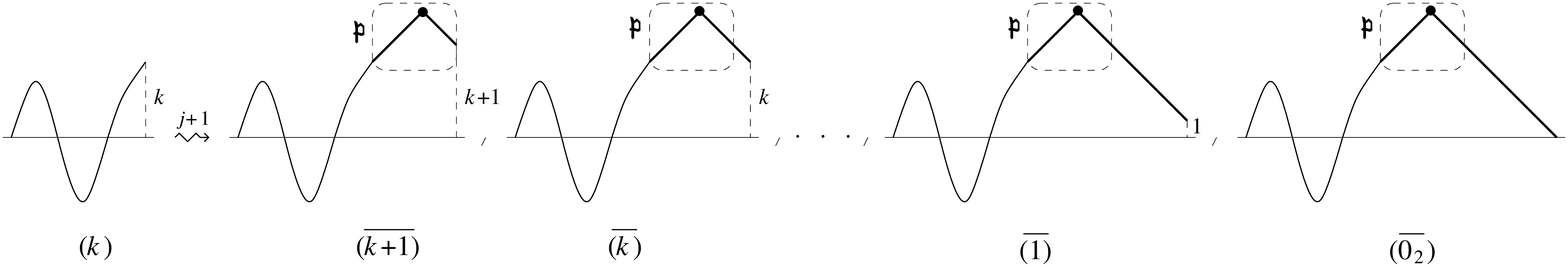,width=6.2in,clip=}
\caption{\small{The mapping associated to $(k)
\stackrel{j+1}{\rightsquigarrow}
(\overline{0_2})(\overline{1})\ldots(\overline{k+1})$ of
(\ref{Rule})} \label{soprasegnato}}\vspace{-15pt}
\end{center}
\end{figure}

A lattice path $\omega \in F$, with $n$ rise steps and such that
its endpoint has ordinate $k$, provides $k+3$ lattice paths, with
$n+j+1$ rise steps, according to the second production of
(\ref{Rule}) having $0,0,1,\ldots,k+1$ as endpoint ordinate,
respectively. The last $k+2$ labels are obtained by adding to
$\omega$ a sequence of steps consisting of the marked forbidden
pattern $\mathfrak{p}=x^{j+1}\overline{x}^j$ followed by $k+1-h$,
$0 \leq h \leq k+1$, fall steps (see Figure \ref{soprasegnato}).
Each lattice path so obtained has the property that its rightmost
suffix beginning from the $x$-axis, either remains strictly above
the $x$-axis itself or ends on the $x$-axis by a fall step. At
this point the label $(0_1)$ due to the first and the second
production of (\ref{Rule}) yield lattice paths which either do not
contain marked forbidden patterns in its rightmost suffix and end
on the $x$-axis by a rise step or having the rightmost marked
point with ordinate less than or equal to $j$.

\begin{figure}[!htb]
\begin{center}
\epsfig{file=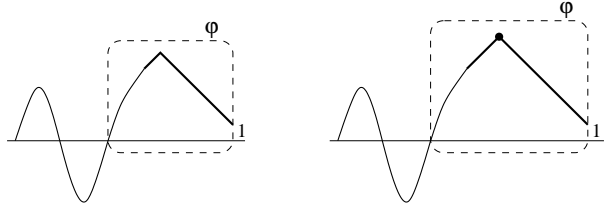,width=3.1in,clip=} \caption{\small{A graphical
representation of the suffix $\varphi$ in $\omega'$}
\label{fifi}}\vspace{-15pt}
\end{center}
\end{figure}

In order to obtain the label $(0_1)$ according to the first
production of (\ref{Rule}), we consider the lattice path $\omega'$
obtained from $\omega$ by adding a sequence of steps consisting of
one rise step followed by $k$ fall steps, while in order to obtain
the label $(0_1)$ according to the second production of
(\ref{Rule}), we consider the lattice path $\omega'$ obtained from
$\omega$ by adding a sequence of steps consisting of the marked
forbidden pattern $\mathfrak{p}=x^{j+1}\overline{x}^j$ followed by
$k$ fall steps. By applying the previous actions, a path $w'$ can
be written as $w'= v \varphi$, where $\varphi$ is the rightmost
suffix in $w'$ beginning from the $x$-axis and strictly remaining
above the $x$-axis (see Figure \ref{fifi}).

We distinguish two cases: in the first one $\varphi$ does not
contain any marked point and in the second one $\varphi$ contains
at least one marked point.

\begin{figure}[htb]
\begin{center}
\epsfig{file=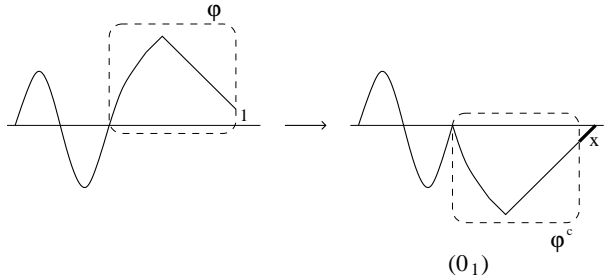,width=3.1in,clip=}
\caption{\small{A graphical representation of the actions giving
the label $(0_1)$ in case of no marked points in $\varphi$}
\label{semplice}}\vspace{-15pt}
\end{center}
\end{figure}

If the suffix $\varphi$ does not contain any marked point, then
the desired label $(0_1)$ is associated to the path $v\varphi^cx$,
where ${\varphi}^c$ is the path obtained from $\varphi$ by
switching rise and fall steps (see Figure \ref{semplice}).

\begin{figure}[!htb]
\begin{center}
\epsfig{file=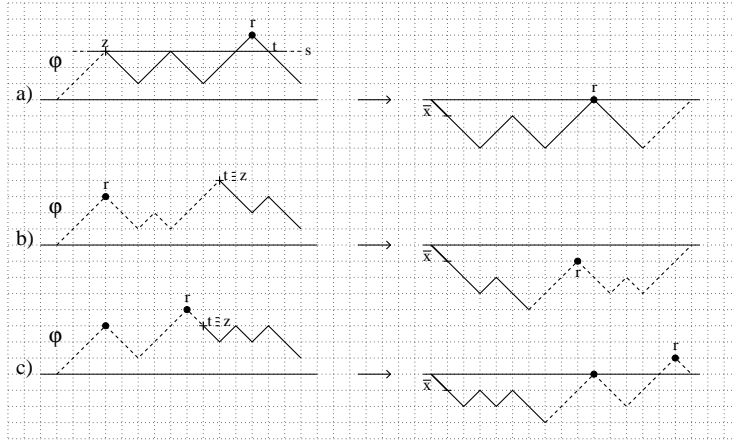,width=3.8in,clip=} \caption{\small{Some
examples of the actions giving the label $(0_1)$ in the case
of marked points in $\varphi$, $\mathfrak{p}=x^2\overline{x}$}
\label{ex}}\vspace{-15pt}
\end{center}
\end{figure}

If the suffix $\varphi$ contains marked points, let $r$ be the
rightmost marked point in $\varphi$ having highest ordinate and
$t$ be the nearest point on the right of the marked forbidden
pattern containing $r$ with highest ordinate and which is not
strictly within a marked forbidden pattern. We consider the
straight line $s$ through the point $t$ and the leftmost point $z$
in $\varphi$ with highest ordinate, which lies above or on the
line $s$ and which is not strictly within a marked forbidden
pattern (see the left side of Figure \ref{ex}.a)). Obviously, if
the straight line $s$ does not intersect any points on the left of
$t$ (see the left side of Figure \ref{ex}.b)) or intersects only
points lying strictly within a marked forbidden pattern (see the
left side of Figure \ref{ex}.c)), then $z\equiv t$.

\begin{figure}[!htb]
\begin{center}
\epsfig{file=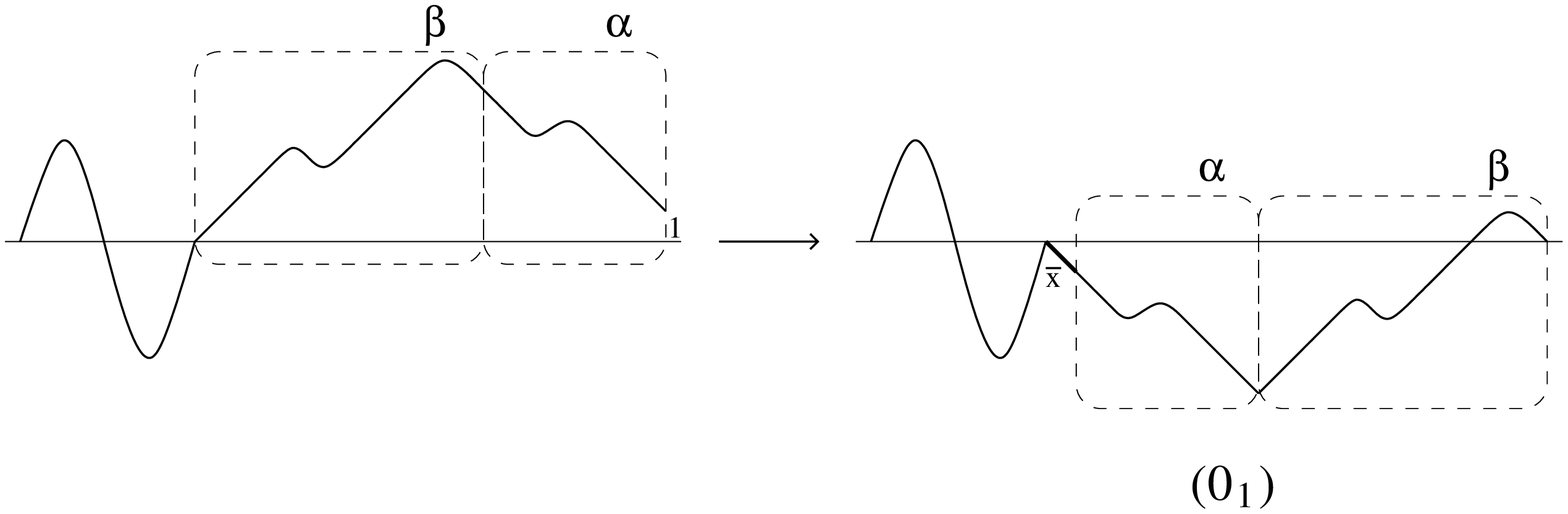,width=4.2in,clip=}
\caption{\small{A graphical representation of the cut and paste
actions giving the label $(0_1)$ in case of marked points in
$\varphi$}\label{ardita}}\vspace{-15pt}
\end{center}
\end{figure}

The desired label $(0_1)$ is associated to the path
obtained by concatenating a fall step $\overline{x}$ with the path
in $\varphi$ running from $z$ to the endpoint of the path, say
$\alpha$, and the path running
from the initial point in $\varphi$ to $z$, say $\beta$ (see Figure \ref{ex} and \ref{ardita}).

\begin{figure}[!htb]
\begin{center}
\epsfig{file=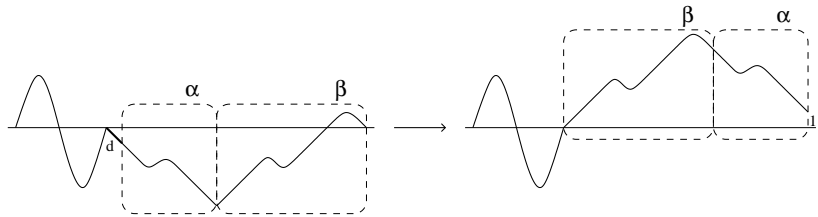,width=4.2in,clip=} \caption{\small{A
graphical representation of the lattice path obtained by means of
the inverted mapping related to the label $(0_1)$ in case of
marked points in $\varphi$}\label{inve}}\vspace{-15pt}
\end{center}
\end{figure}

This last mapping can be inverted as follows. Let $d$ be the
rightmost fall step in a path $\omega'$ labelled $(0_1)$ such that
it begins from the $x$-axis and each marked point, on its right,
has ordinate less than or equal to $j$. Let $\omega'=\omega d
\varphi'$ and $l$ the rightmost point in $\varphi'$ with lowest
ordinate. The inverted lattice path of $\omega'$ is given by
$\omega \beta \alpha$, where $\beta$ is the path in $\varphi'$
running from $l$ to the endpoint of the path and $\alpha$ is the
path running from the initial point in $\varphi'$ to $l$ (see
Figure \ref{inve}).

\begin{figure}[!htb]
\begin{center}
\epsfig{file=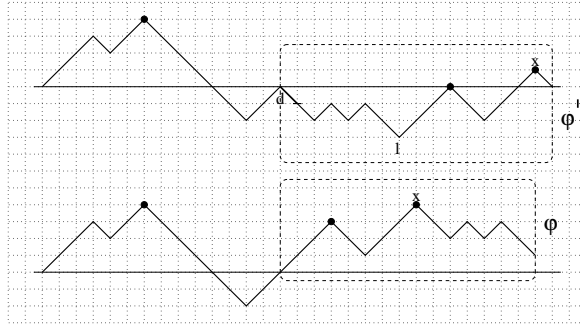,width=3in,clip=}
\caption{\small{The inverted mapping related to the label
$(0_1)$ in case of marked points in $\varphi$}
\label{invertmap}}
\end{center}
\end{figure}

Figure \ref{invertmap} shows the cut and paste actions
related to the inverted mapping with the pattern
$\mathfrak{p}=x^2\overline{x}$.

\begin{figure}[!htb]
\begin{center}
\epsfig{file=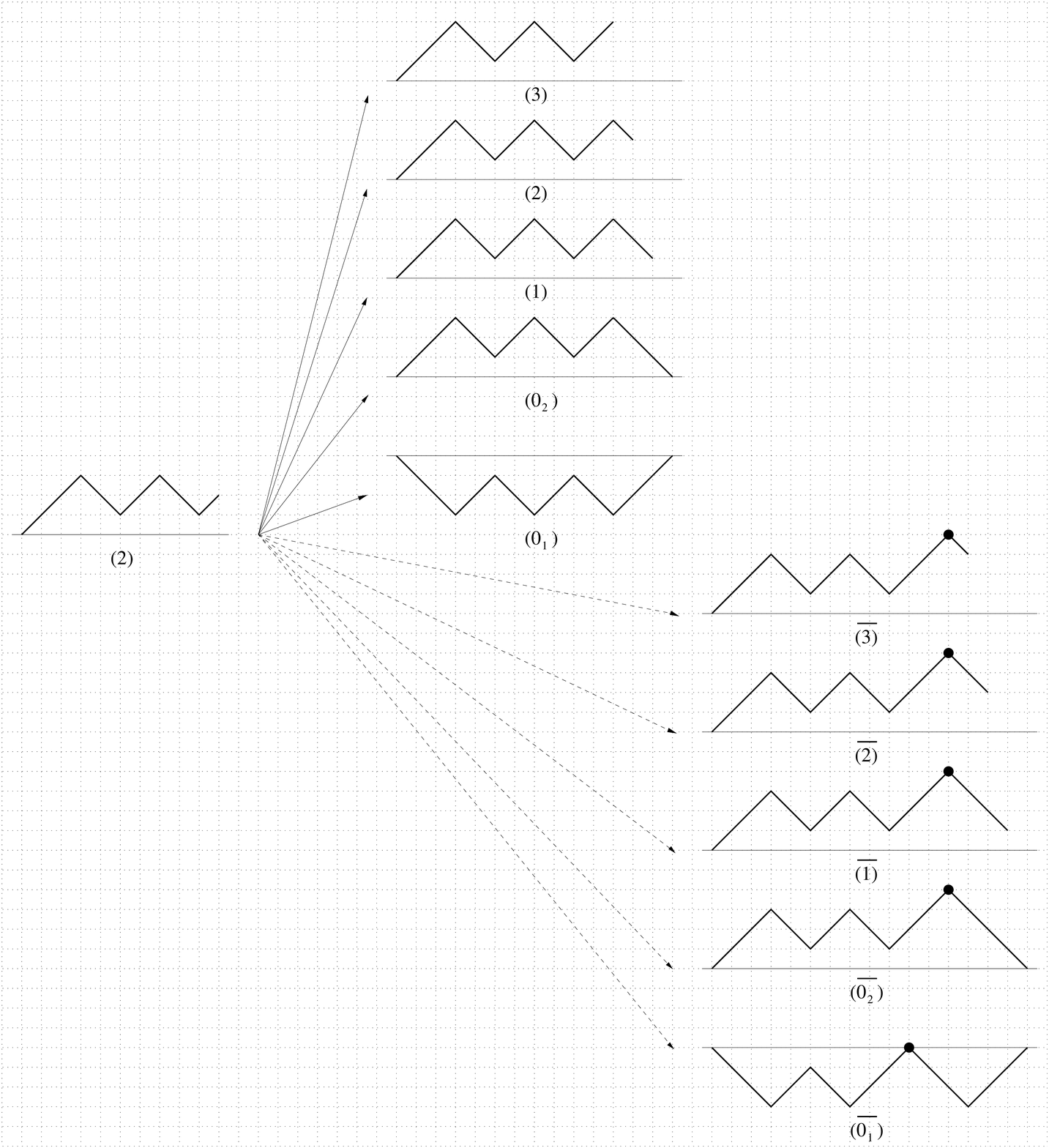,width=4in,clip=}
\caption{\small{The set of lattice paths obtained from a given
$(k)$, by means of the succession rule (\ref{Rule})}
\label{esempione}}\vspace{-15pt}
\end{center}
\end{figure}

\begin{figure}[!htb]
\begin{center}
\epsfig{file=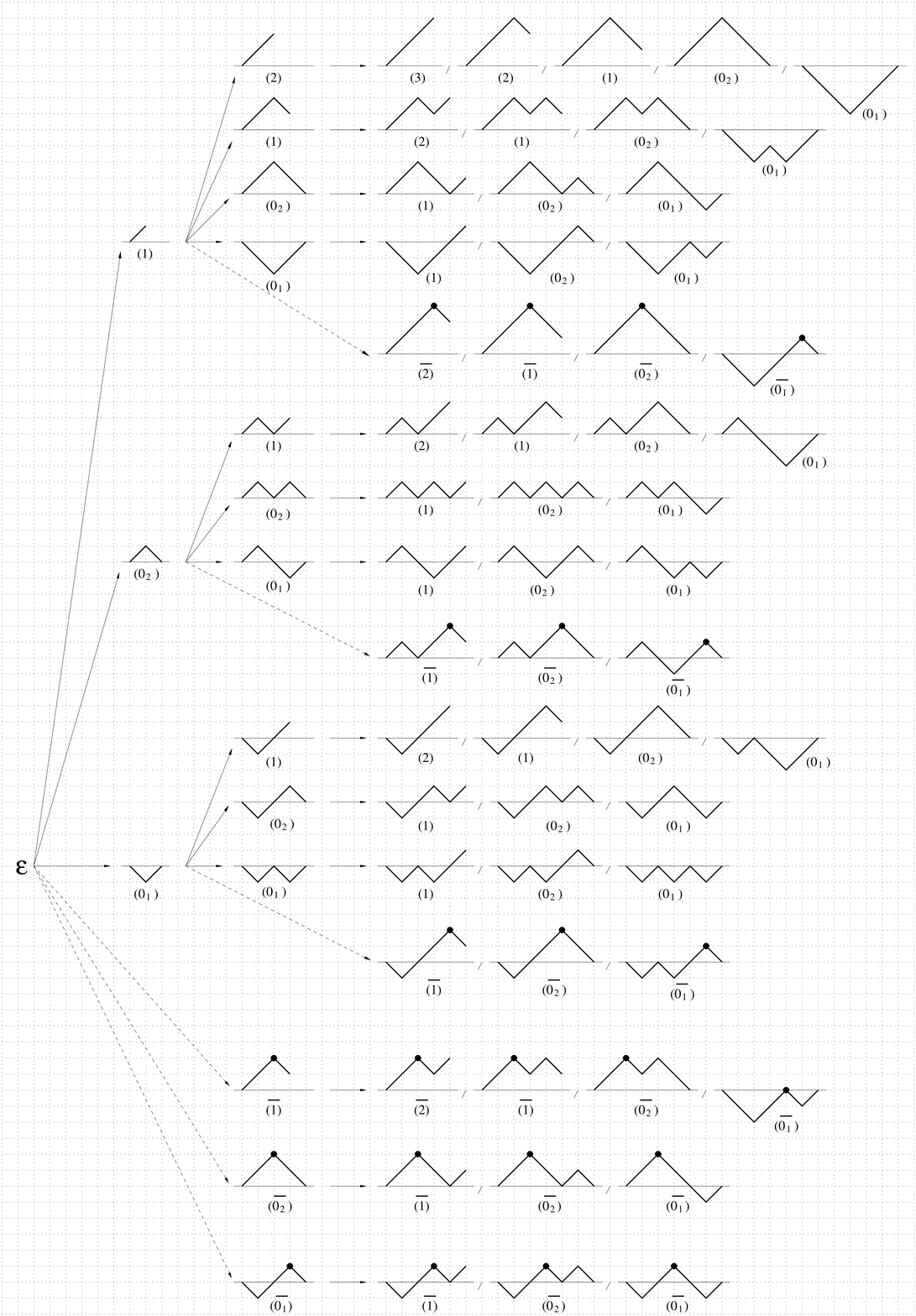,width=4.7in,clip=} \caption{\small{Some
levels of the generating tree associated with the succession rule
(\ref{Rule}) for the path in $F^{[\mathfrak{p}]}$, being
$\mathfrak{p}=x^2\overline{x}$} \label{alberino}}
\end{center}
\end{figure}

At this point, we can describe the complete mapping
defined by the succession rule (\ref{Rule}). In particular Figure
\ref{esempione} shows this complete mapping with the pattern
$\mathfrak{p}=x^2\overline{x}$ and Figure \ref{alberino} sketches
some levels of the generating tree for the paths in
$F^{[\mathfrak{p}]}$ enumerated according to the number of the
rise steps.

\begin{figure}[!htb]
\begin{center}
\epsfig{file=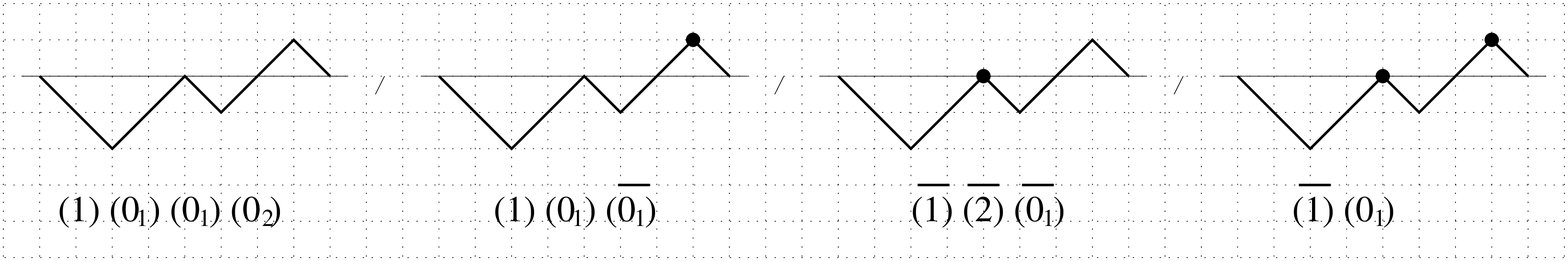,width=4.75in,clip=}
\caption{\small{The 4 copies of a given path having 2 forbidden
patterns, $\mathfrak{p}=x^2\overline{x}$}
\label{pardis}}\vspace{-15pt}
\end{center}
\end{figure}

This construction generates $2^C$ copies of each path having $C$
forbidden patterns such that $2^{C-1}$ instances are coded by a
sequence of labels ending by a marked one, say $(\overline{k})$,
and contain an odd number of marked forbidden patterns, and
$2^{C-1}$ instances are coded by a sequence of labels ending by a
non-marked one, say $(k)$, and contain an even number of marked
forbidden patterns. For example, Figure \ref{pardis} shows the 4
copies of a given path having 2 forbidden patterns
$\mathfrak{p}=x^2\overline{x}$, where the sequences of labels show
the derivation of each path in the generating tree.

This observation is due to the fact that when a path is obtained
according to the first production of (\ref{Rule}) then no marked
forbidden pattern is added. Moreover, when a path is obtained
according to the second production of (\ref{Rule}) exactly one
marked forbidden pattern is added. In any case, the actions
performed to obtain the label $(0_1)$ do not change the number
of marked forbidden patterns in the path.

\begin{theorem} The generating tree of the lattice paths in $F^{[\mathfrak{p}]}$, where
$\mathfrak{p}=x^{j+1}\overline{x}^j$, according to the number of
rise steps, is isomorphic to the tree having its root labelled
$(0)$ and recursively defined by the succession rule
$(\ref{Rule})$.
\end{theorem}

\emph{Proof.}\quad We have to show that the algorithm described in
the previous pages is a construction for the set
$F^{[\mathfrak{p}]}$ according to the number of rise steps. This
means that all the paths in $F$ with $n$ rise steps are obtained.
Moreover, for each obtained path $\omega$ in $F \backslash
F^{[\mathfrak{p}]}$, having $C$ forbidden patterns, with $n$ rise
steps and $(k)$ as last label of the associated code, a path
$\omega'$ in $F \backslash F^{[\mathfrak{p}]}$ with $n$ rise
steps, $C$ forbidden patterns and $(\overline{k})$ as last label
of the associated code is also generated having the same form as
$\omega$ but such that the last forbidden pattern is marked if it
is not in $\omega$
and vice-versa.

The first assertion is an immediate consequence of the construction
according to the first production of (\ref{Rule}).

In order to prove the second assertion we have to distinguish two
cases (which in their turn are subdivided in 5 and 3 subitems
respectively) depending on whether the last forbidden pattern is
marked or not. For sake of completeness we report the entire
proof, which is indeed rather cumbersome. Anyhow, the interested
reader could skip all the subitems, except the first ones. In
fact, all the others are obtained from these by means of slight
modifications.

We denote by $h$ be the ordinate of the peak of the last forbidden
pattern.\\

First case: the last forbidden pattern in $\omega$ is marked.
We consider the following subcases:
$h>j$, $h=j$, $0<h<j$, $h=0$ and $h<0$.
\begin{itemize}
\item[1) $h>j$:] Each path $\omega$ in $F \backslash
F^{[\mathfrak{p}]}$ can be written as $\omega=\mu x^{j+1}
{\overline{x}}^{f} \nu$, where $\mu \in F$, $\nu \in
F^{[\mathfrak{p}]}$ and $j \leq f \leq d+j+1$ where $d \geq 0$ is
the ordinate of the endpoint of $\mu$ (see Figure \ref{maggiore}).

\begin{figure}[!htb]
\begin{center}
\epsfig{file=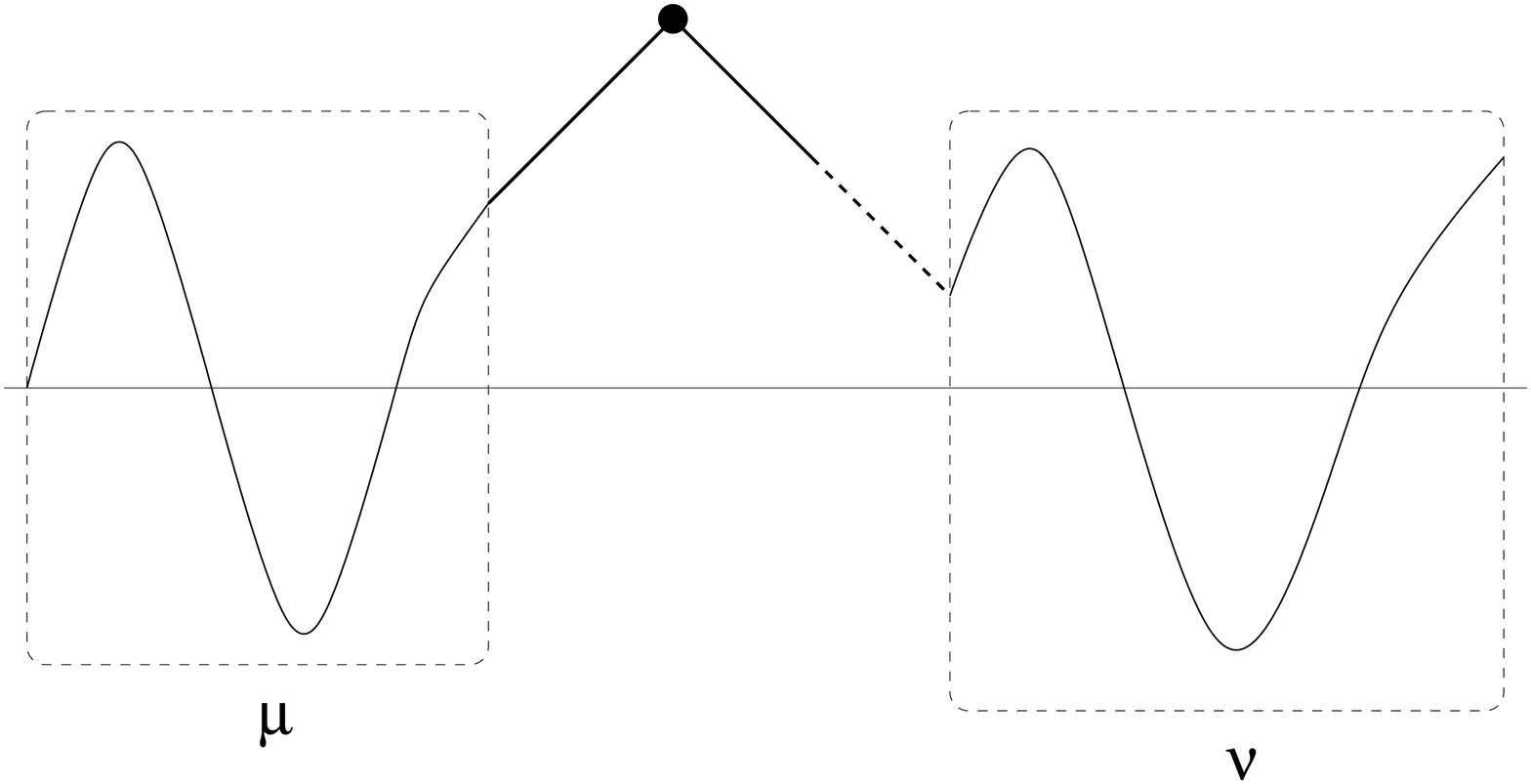,width=2.8in,clip=} \caption{\small{A
graphical representation of the path $\omega$ in the case $h>j$}
\label{maggiore}}\vspace{-15pt}
\end{center}
\end{figure}

The path $\omega'$ which kills $\omega$ is obtained by performing
on $\mu$ the following: add the path $x^j$ by applying $j$ times
the mapping associated to $(k)\stackrel{1}{\rightsquigarrow}(k+1)$
of the first production of (\ref{Rule}), add the path $x
{\overline{x}}^{f}$ by applying the mapping associated to
$(k)\stackrel{1}{\rightsquigarrow}(d+j+1-f)$ of the first
production of (\ref{Rule}). The path $\nu$ in $\omega'$ is
obtained as in $\omega$. \item[2) $h=j$:] Each path $\omega$ in $F
\backslash F^{[\mathfrak{p}]}$ can be written as $\omega=\mu
\overline{x} \gamma x^{j+1} {\overline{x}}^{j} \nu$, where
$\mu,\gamma \in F$ and $\nu \in F^{[\mathfrak{p}]}$ (see Figure
\ref{uguale}). We observe that the path $\gamma$ can contain
marked points, with ordinate $b < j$, or not. If the path $\gamma$
contains no marked point, then it remains strictly under the
$x$-axis, otherwise the marked forbidden patterns intersect the
$x$-axis when $0\leq b < j$. In the following cases we consider a
path $\gamma$ having the same property.

\begin{figure}[!htb]
\begin{center}
\epsfig{file=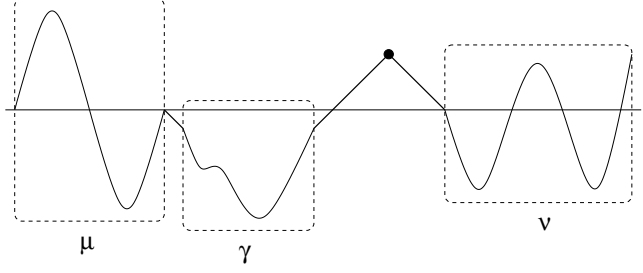,width=3.3in,clip=} \caption{\small{A
graphical representation of the path $\omega$ in the case $h=j$}
\label{uguale}}\vspace{-15pt}
\end{center}
\end{figure}

The path $\omega'$ which kills $\omega$ is obtained by performing
on $\mu \overline{x} \gamma x$ the following: add the path
$x^{j-1}$ by applying $j-1$ times the mapping associated to
$(k)\stackrel{1}{\rightsquigarrow}(k+1)$ of the first production
of (\ref{Rule}), add the path $x {\overline{x}}^{j}$ by applying
the mapping associated to $(k)\stackrel{1}{\rightsquigarrow}(0_2)$
of the first production of (\ref{Rule}).
The path $\nu$ in $\omega'$ is obtained as in $\omega$.
\item[3) $0<h<j$:] Each path $\omega$ in $F \backslash
F^{[\mathfrak{p}]}$ can be written as $\omega=\mu \overline{x}
\gamma x^{j+1} {\overline{x}}^{j}\eta x \nu$, where $\mu,\gamma
\in F$ and $\eta,\nu \in F^{[\mathfrak{p}]}$ (see Figure
\ref{compreso}). We observe that the path $\eta$ remains strictly
under the $x$-axis. In the following cases we consider a path
$\eta$ having the same property.

\begin{figure}[!htb]
\begin{center}
\epsfig{file=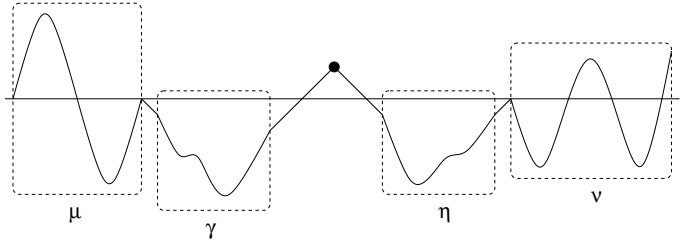,width=3.5in,clip=} \caption{\small{A
graphical representation of the path $\omega$ in the case $0<h<j$}
\label{compreso}}\vspace{-15pt}
\end{center}
\end{figure}

The path $\omega'$ which kills $\omega$ is obtained by performing
on $\mu \overline{x} \gamma x^{j+1-h}$ the following: add the path
$x^{h-1}$ by applying $h-1$ times the mapping associated to
$(k)\stackrel{1}{\rightsquigarrow}(k+1)$ of the first production
of (\ref{Rule}), add the path $x {\overline{x}}^{h}$ by applying
the mapping associated to $(k)\stackrel{1}{\rightsquigarrow}(0_2)$
of the first production of (\ref{Rule}), add
the path ${\overline{x}}^{j-h} \eta x$ by applying consecutive and
appropriate mappings of the first production of (\ref{Rule}) and these
mappings must be completed by performing the actions giving the label $(0_1)$
in case of no marked points. The path $\nu$ in
$\omega'$ is obtained as in $\omega$. \item[4) $h=0$:] Each path
$\omega$ in $F \backslash F^{[\mathfrak{p}]}$ can be written as
$\omega=\mu \overline{x} \gamma x^{j+1} {\overline{x}}^{j} \eta x
\nu$, where $\mu,\gamma \in F$ and $\eta,\nu \in
F^{[\mathfrak{p}]}$ (see Figure \ref{ugualeazero}).

\begin{figure}[!htb]
\begin{center}
\epsfig{file=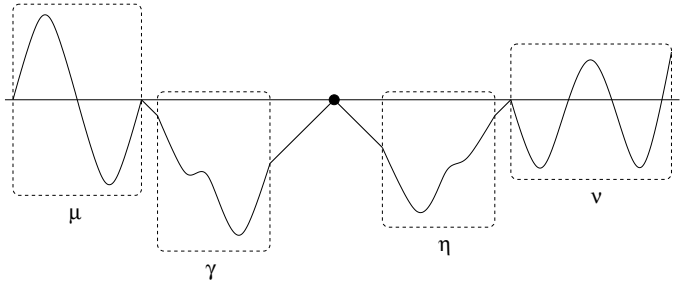,width=3.5in,clip=} \caption{\small{A
graphical representation of the path $\omega$ in the case $h=0$}
\label{ugualeazero}}\vspace{-15pt}
\end{center}
\end{figure}

The path $\omega'$ which kills $\omega$ is obtained by performing
on $\mu \overline{x} \gamma x^{j+1}$ the following: add the path
${\overline{x}}^{j} \eta x$ by applying consecutive and
appropriate mappings of the first production of (\ref{Rule}),
apply the actions giving the label $(0_1)$ in case of no
marked points. The path $\nu$ in $\omega'$ is obtained as in
$\omega$. \item[5) $h<0$:] Each path $\omega$ in $F \backslash
F^{[\mathfrak{p}]}$ can be written as $\omega=\mu \overline{x}
\gamma x^{j+1} {\overline{x}}^{j} \eta x \nu$, where $\mu,\gamma
\in F$ and $\eta,\nu \in F^{[\mathfrak{p}]}$ (see Figure
\ref{minore}).

\begin{figure}[!htb]
\begin{center}
\epsfig{file=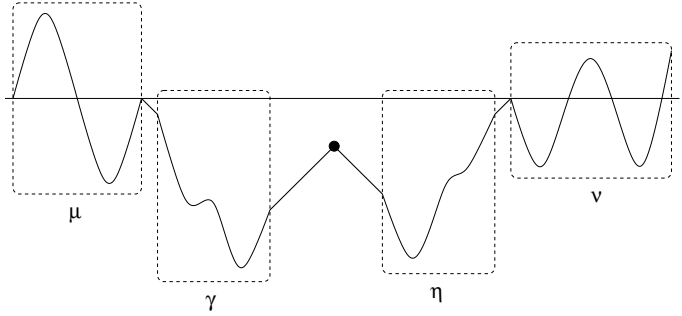,width=3.5in,clip=} \caption{\small{A
graphical representation of the path $\omega$ in the case $h<0$}
\label{minore}}\vspace{-15pt}
\end{center}
\end{figure}

We distinguish two subcases: in the first one the path $\gamma$
contains no marked points and remains strictly under the $x$-axis
and in the
second one the path $\gamma$ contains at least a marked point.

In the first subcase, the path $\omega'$ which kills $\omega$ is
obtained by performing on $\mu$ the following: add the path
$\overline{x} \gamma x^{j+1} {\overline{x}}^{j} \eta x$ by
applying consecutive and appropriate mappings of the first
production of (\ref{Rule}), apply the actions giving the
label $(0_1)$ in case of no marked points. The path
$\nu$ in $\omega'$ is obtained as in $\omega$.

In the second subcase, we consider the rightmost point $l$ of the
path $\overline{x} \gamma x^{j+1} {\overline{x}}^{j} \eta x$ with
lowest ordinate. The path $\omega'$ which kills $\omega$ is
obtained by performing on $\mu$ the following: add the path in
$\gamma x^{j+1} {\overline{x}}^{j}\eta x$ running from $l$ to the
endpoint of the path by applying consecutive and appropriate
mappings of the first and second production of (\ref{Rule}), add
the path in $\gamma x^{j+1} {\overline{x}}^{j}\eta x$ running from
its initial point to $l$ by applying consecutive and appropriate
mappings of the first and second production of (\ref{Rule}), apply
the cut and paste actions giving the label $(0_1)$ in case of
marked points. Obviously, the last forbidden pattern in the path
must be generated by applying consecutive and appropriate mappings
of the first production of (\ref{Rule}). The path $\nu$ in
$\omega'$ is obtained as in $\omega$.
\end{itemize}

Second case: the last forbidden pattern in $\omega$ is not a
marked forbidden pattern.
We consider the following subcases:
$h>j$, $h=j$ and $h < j$.
\begin{itemize}
\item[1) $h>j$:] Each path $\omega$ in $F \backslash
F^{[\mathfrak{p}]}$ can be written as $\omega=\mu x^{j+1}
{\overline{x}}^{f} \nu$, where $\mu \in F$, $\nu \in
F^{[\mathfrak{p}]}$ and $j \leq f \leq d+j+1$ where $d \geq 0$ is
the ordinate of the endpoint of $\mu$ (Figure
\ref{maggiorenoseg}).

\begin{figure}[!htb]
\begin{center}
\epsfig{file=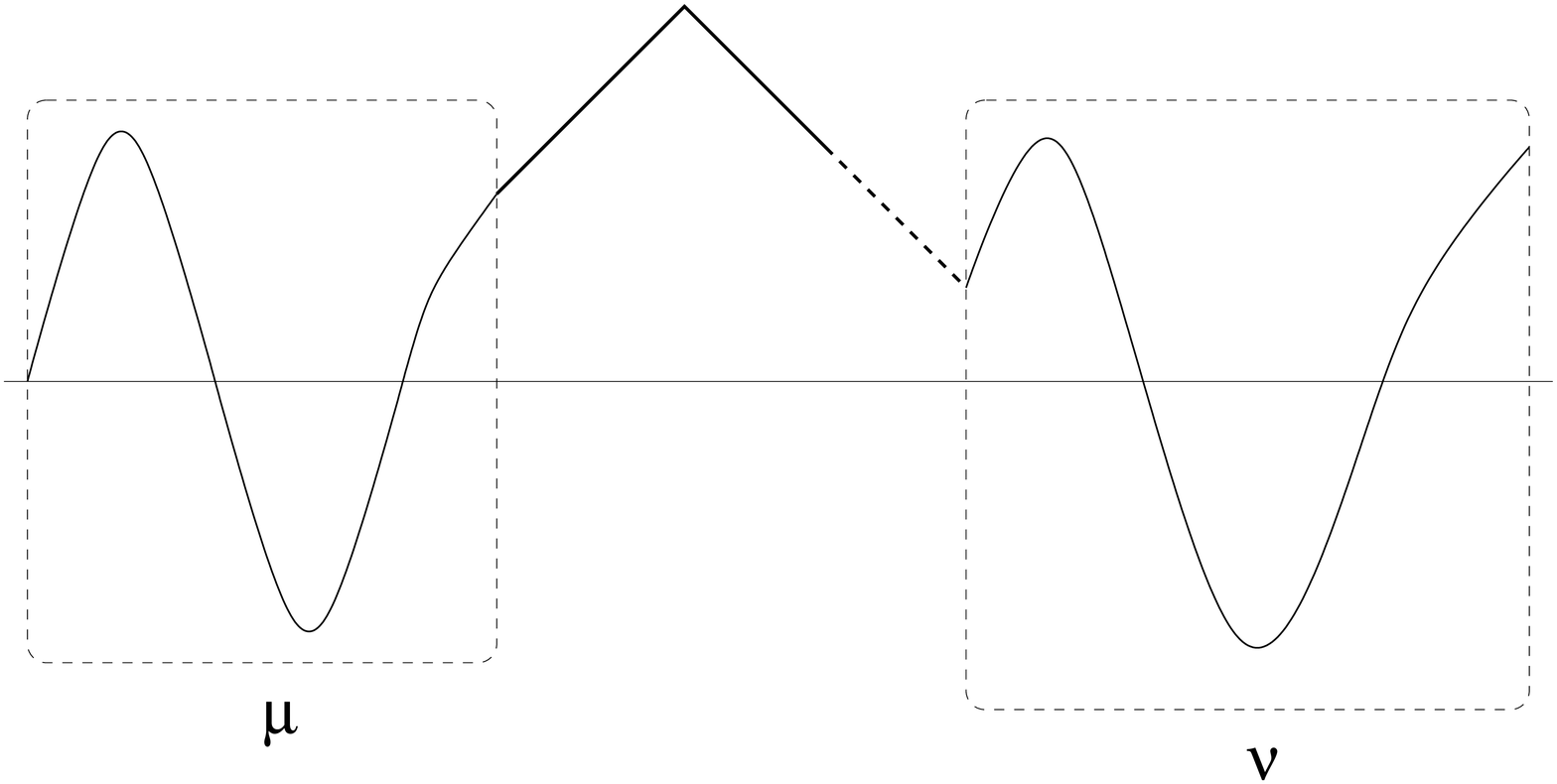,width=2.8in,clip=} \caption{\small{A
graphical representation of the path $\omega$ in the case $h>j$}
\label{maggiorenoseg}}\vspace{-15pt}
\end{center}
\end{figure}

The path $\omega'$ which kills $\omega$ is obtained by performing
on $\mu$ the following: add the path $x^{j+1}{\overline{x}}^f$ by
applying the mapping associated to
$(k)\stackrel{j+1}{\rightsquigarrow}(\overline{d+j+1-f})$ of the
second production of (\ref{Rule}). The path $\nu$ in $\omega'$ is
obtained as in $\omega$. \item[2) $h=j$:] Each path $\omega$ in $F
\backslash F^{[\mathfrak{p}]}$ can be written as $\omega=\mu
\overline{x} \gamma x^{j+1} {\overline{x}}^{j} \nu$, where
$\mu,\gamma \in F$ and $\nu \in F^{[\mathfrak{p}]}$ (see Figure
\ref{ugualesecond}). We observe that the path $\gamma$ can
contains marked points, with ordinate $b < j$, or not. If the path
$\gamma$ contains no marked point, then it remains strictly under
the $x$-axis, otherwise the marked forbidden patterns intersect
the $x$-axis when $0\leq b < j$. In the following case we consider
a path $\gamma$ having the same property.

\begin{figure}[!htb]
\begin{center}
\epsfig{file=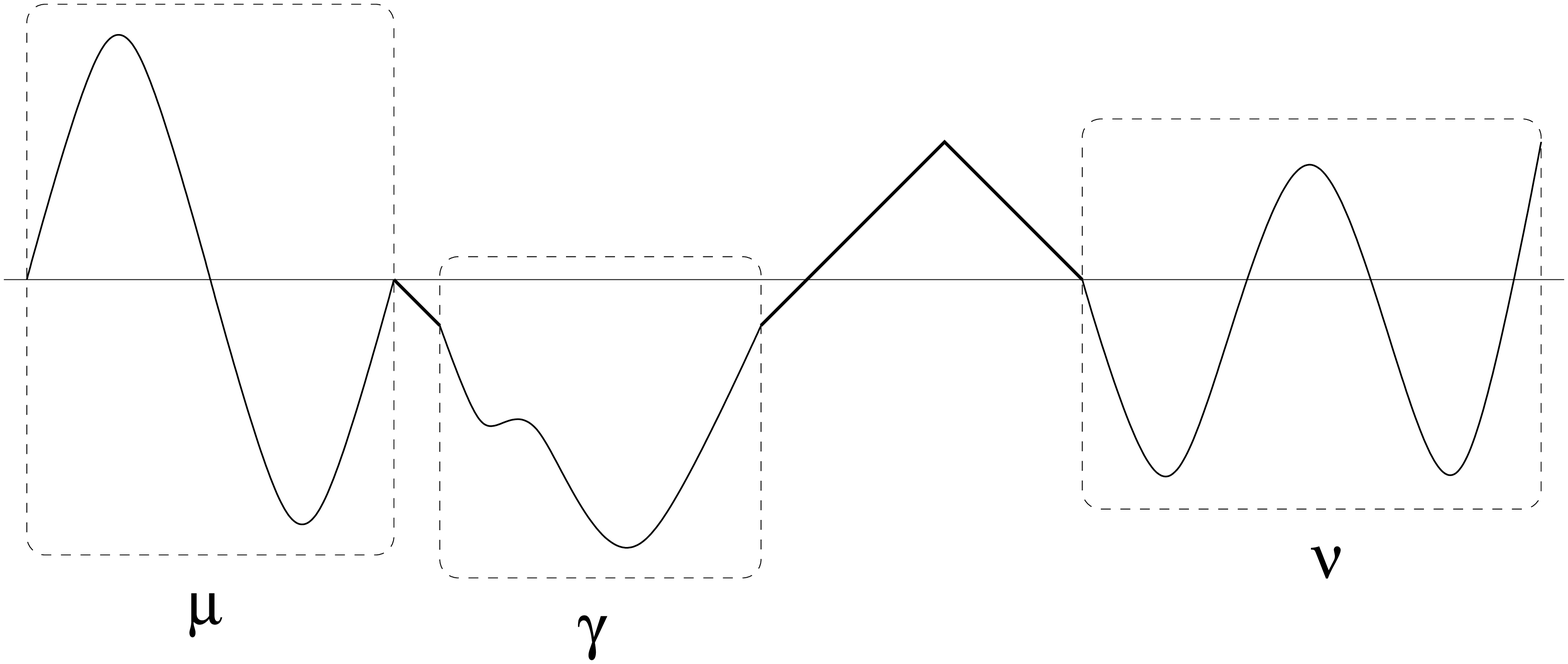,width=3.3in,clip=} \caption{\small{A
graphical representation of the path $\omega$ in the case $h=j$}
\label{ugualesecond}}\vspace{-15pt}
\end{center}
\end{figure}

Let $l$ be the rightmost point of the path $\overline{x} \gamma
x^{j+1} {\overline{x}}^{j}$ with lowest ordinate. The path
$\omega'$ which kills $\omega$ is obtained by performing on $\mu$
the following: add the path in $\gamma x^{j+1} {\overline{x}}^{j}$
running from $l$ to the endpoint of the path by applying
consecutive and appropriate mappings of the first and second
production of (\ref{Rule}), add the path in $\gamma x^{j+1}
{\overline{x}}^{j}$ running from its initial point to $l$ by
applying consecutive and appropriate mappings of the first and
second production of (\ref{Rule}), apply the cut and paste actions
giving the label $(0_1)$ in case of marked points. Obviously,
the last forbidden pattern in the path must be generated by
applying the mapping of the second production of (\ref{Rule}). The
path $\nu$ in $\omega'$ is obtained as in $\omega$. \item[3) $h <
j$:] Each path $\omega$ in $F \backslash F^{[\mathfrak{p}]}$ can
be written as $\omega=\mu \overline{x} \gamma x^{j+1}
{\overline{x}}^{j} \eta x \nu$, where $\mu,\gamma \in F$ and
$\eta,\nu \in F^{[\mathfrak{p}]}$ (see Figure \ref{minorenoseg}).
We observe that the path $\eta$ remains strictly under the
$x$-axis.

\begin{figure}[!htb]
\begin{center}
\epsfig{file=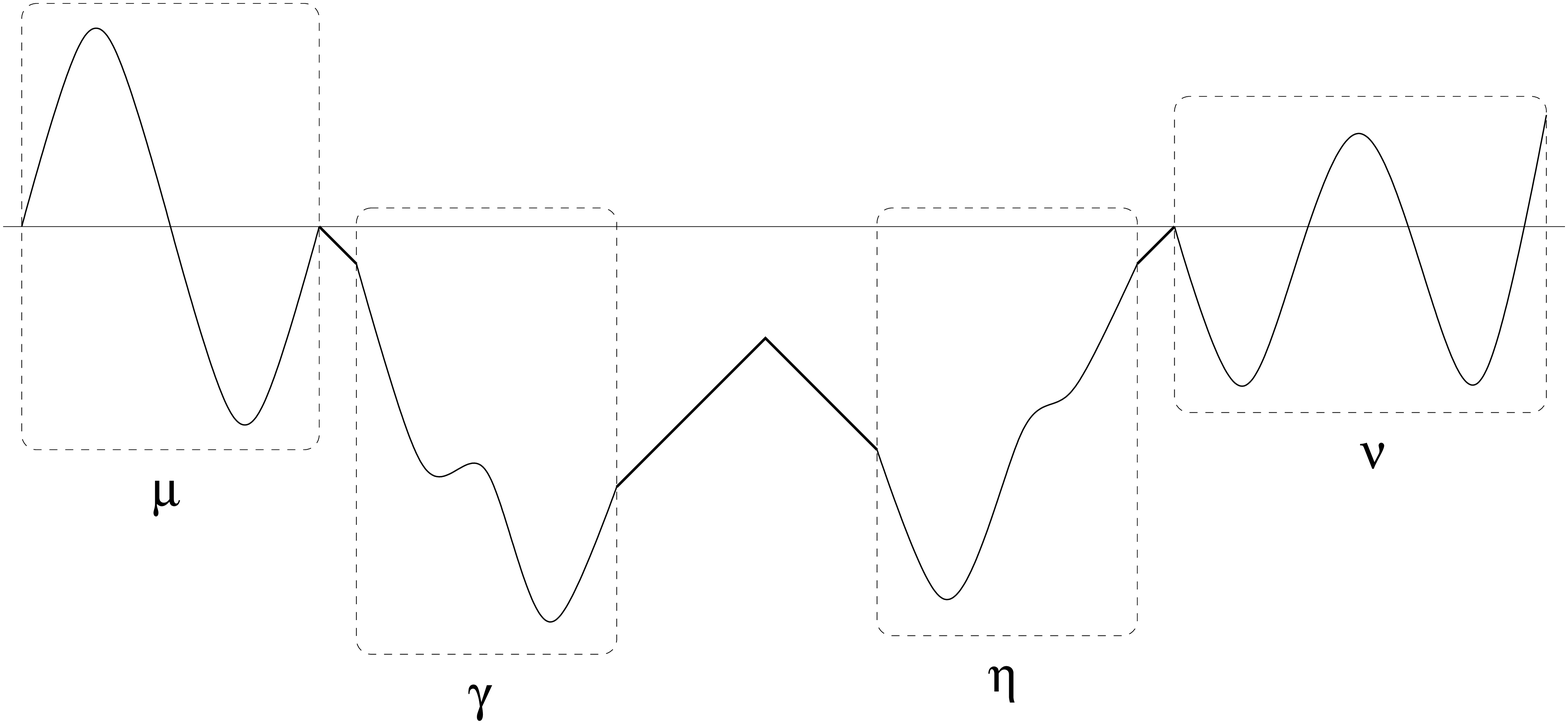,width=3.5in,clip=} \caption{\small{A
graphical representation of the path $\omega$ in the case $h < j$}
\label{minorenoseg}}\vspace{-15pt}
\end{center}
\end{figure}

Let $l$ be the rightmost point of the path $\overline{x} \gamma
x^{j+1} {\overline{x}}^{j}\eta x$ with lowest ordinate. The path
$\omega'$ which kills $\omega$ is obtained by performing on $\mu$
the following: add the path in $\gamma x^{j+1}
{\overline{x}}^{j}\eta x$ running from $l$ to the endpoint of the
path by applying consecutive and appropriate mappings of the first
and second production of (\ref{Rule}), add the path in $\gamma
x^{j+1} {\overline{x}}^{j}\eta x$ running from its initial point
to $l$ by applying consecutive and appropriate mappings of the
first and second production of (\ref{Rule}), apply the cut and
paste actions giving the label $(0_1)$ in case of marked
points. Obviously, the last forbidden pattern in the path must be
generated by applying the mapping of the second production of
(\ref{Rule}). The path $\nu$ in $\omega'$ is obtained as in
$\omega$.
\end{itemize}

We observe that for each path $\omega$ in $F \backslash
F^{[\mathfrak{p}]}$, having $C$ forbidden patterns, with $n$ rise
steps and last label $(k)$, there exists one and only one path
$\omega'$ in $F \backslash F^{[\mathfrak{p}]}$ with $n$ rise
steps, $C$ forbidden patterns and last label $(\overline{k})$
having the same form as $\omega$ but such that the last forbidden
pattern is marked if it is not in $\omega$ and vice-versa.

This assertion is an immediate consequence of the constructions in
the proof, since the described actions are univocally determined.
Therefore, it is not possible to obtain a path $\omega'$ which
kills a given path $\omega$ applying two distinct procedures. \cvd

\section{Conclusions and further developments}
In this paper we study the enumeration, according to the number of
ones, of particular binary words excluding a fixed pattern
$\mathfrak{p}=1^{j+1}0^j$, $j \geq 1$. Initially, we have solved
the problem algebraically by means of Riordan arrays. This
approach allows us to obtain a jumping and marked succession rule
describing the growth of such words. Note that, it is not
possible to associate to a word a path in the generating tree
obtained by the succession rule. This problem is solved by means
of an algorithm constructing all binary words having a fixed
number of ones and eliminating the words which contain the
forbidden pattern $\mathfrak{p}=1^{j+1}0^j$, $j \geq 1$.

Further developments could investigate for a unified proof simpler
than the one given in this paper. Successive studies should take
into consideration binary words avoiding different forbidden
patterns both from an enumerative and a constructive point of
view. A first step could be the generalization of the forbidden
pattern $\mathfrak{p}$, passing from $\mathfrak{p}=1^{j+1}0^j$, $j
\geq 1$ to $\mathfrak{p}=1^j0^i$, $0<i<j$.

Afterwords, it should be
interesting to study words avoiding patterns having a different
shape, that is not only patterns consisting of a sequence of rise
steps followed by a sequence of fall steps. One could also
consider forbidden patterns on an arbitrary alphabet and to
investigate the properties both of words avoiding that pattern and
of the combinatorial objects.

Finally, we could think of studying words avoiding more than one
pattern and the related combinatorial objects, considering various
parameters.

\section{Acknowledgements}
The authors wish to thank the anonymous reviews whose comments and suggestions
greatly improved the readability of the present paper.

\end{document}